\documentstyle[aasj,epsfig]{aipproc}

\newcommand{\rxte}{{\it RXTE\/}}
\newcommand{\Hz}{{\rm Hz}}
\newcommand{\kHz}{{\rm kHz}}
\newcommand{\Mspyr}{\Msun\ {\rm yr}^{-1}}

\def\lesssim{\mathrel{\hbox{\rlap{\hbox{\lower4pt\hbox{$\sim$}}}\hbox{$<$}}}}
\def\gtrsim{\mathrel{\hbox{\rlap{\hbox{\lower4pt\hbox{$\sim$}}}\hbox{$>$}}}}

\newcommand{\Mdot}{\dot{M}}

\newcommand{\Msun}{M_\odot}

\newcommand{\nuspin}{\nu_s}

\newcommand{\zd}{z_{\rm d}}
\newcommand{\dzd}{\Delta\zd}

\newcommand{\Ef}{E_{\rm F}}

\begin{document}

\title{Gravitational Waves from Low-Mass X-ray Binaries: a Status
Report% 
\thanks{ Much theoretical progress has been made in understanding GW
emission from LMXBs in the six months since the Amaldi conference in
July 1999. Rather than just transcribe the talk given by one of us, we
review the situation as of December 1999.  Because of space
limitations, this review is far from complete.}}
\author{Greg Ushomirsky$^\dagger$, 
	 Lars Bildsten$^\S$,
 	 and Curt Cutler$^\ddagger$} 
\address{$^\dagger$Department of Physics and Department of
			Astronomy,\\ 
	University of California, Berkeley, CA 94720\\ 
	$^\S$Institute for Theoretical Physics and Department of Physics,\\
		University of California, Santa Barbara, CA 93106\\
	$^\ddagger$Max-Planck-Institut fuer Gravitationsphysik, 
	Albert-Einstein-Institut, \\
	Am Muehlenberg 1, D-14476 Golm bei Potsdam, Germany}

\maketitle

\begin{abstract}
We summarize the observations of the spin periods of rapidly accreting
neutron stars.  If gravitational radiation is responsible for
balancing the accretion torque at the observed spin frequencies of
$\approx300$~Hz, then the brightest of these systems make excellent
gravitational wave sources for LIGO-II and beyond. We review the
recent theoretical progress on two mechanisms for gravitational wave
emission: mass quadrupole radiation from deformed neutron star crusts
and current quadrupole radiation from r-mode pulsations in neutron
star cores.
\end{abstract}

\section{Spins of Accreting Neutron Stars}
\label{sec:observations}

Gravitational wave emission from rapidly rotating neutron stars (NS)
has attracted considerable interest in the past several years. In
addition to radiation from the spindown of newborn NSs (see
the review by B.~Owen in this volume), it has long been suspected
\cite{Wagoner84,Thorne87:300yrs} that rapidly accreting NSs,
such as Sco~X-1, may be a promising class of gravitational wave (GW)
emitters.  However, firm observational evidence of fast spins of these
neutron stars had been missing until recently.

\begin{figure}
\begin{center}
\epsfig{file=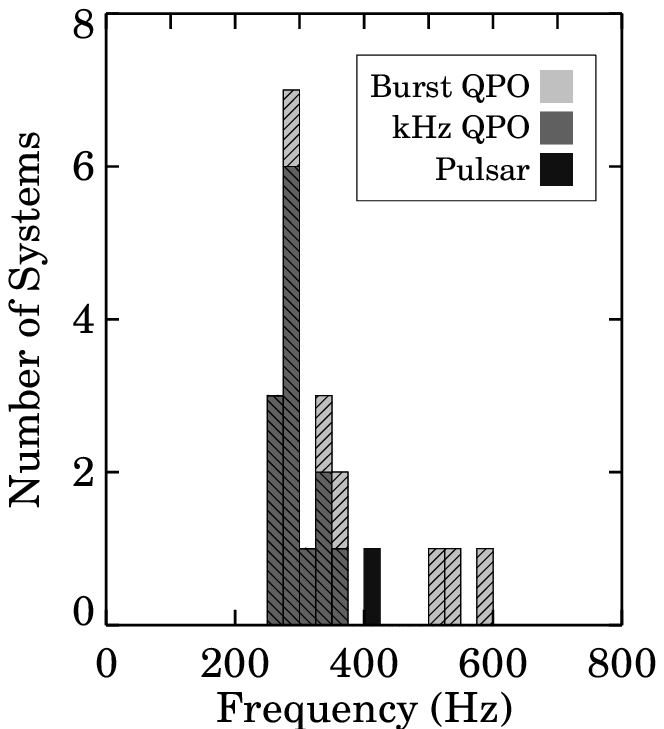,width=2.85in}
\epsfig{file=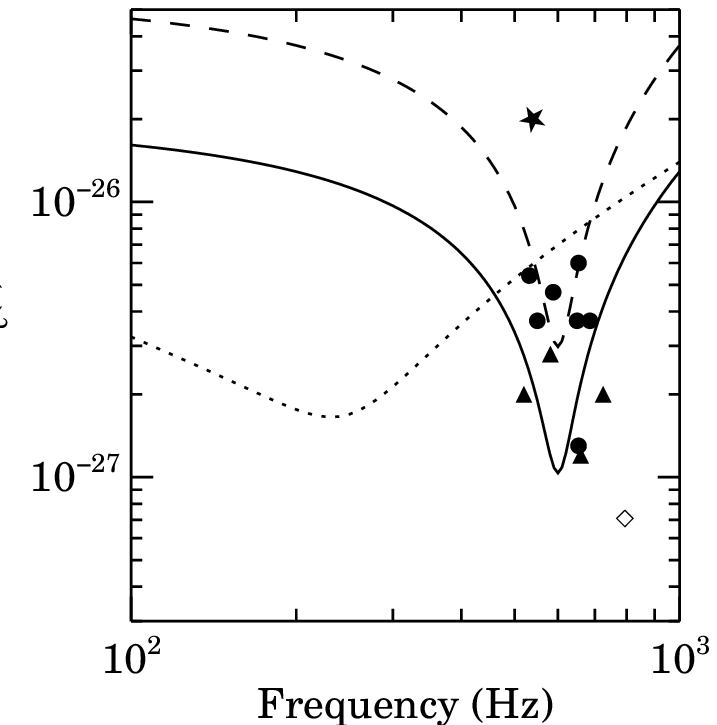, width=2.85in}
\end{center}
\caption[a]{\label{fig:spin-histogram}
Left (a) The distribution of spins of NSs in LMXBs, summarizing
Tables~1-4 of \protect{\cite{klis99:_millis}}.
Right (b)
\label{fig:ligo2} Characteristic signal amplitude $h_c$
for several known LMXBs (symbols, see text) compared with the
sensitivity $h_{3/{\rm yr}}$ of LIGO-II in broadband (the dotted line)
and narrowband (the solid line) configuration (provided by
K. A. Strain on behalf of the LIGO Scientific Community).  The dashed
line shows LIGO-II sensitivity for a two-week integration (see text).}
\end{figure}

NSs in low-mass X-ray binaries (LMXBs) have long been thought to be
the progenitors of millisecond pulsars \cite{bhattacharya95}. However,
directly measuring their periods has proved elusive, probably because
of their rather low magnetic fields.  With the launch of the {\it
Rossi X-ray Timing Explorer}, precision timing of accreting NSs has
opened new threads of inquiry into the behavior and lives of these
objects.  \rxte\ observations \cite{klis99:_millis} have finally
provided conclusive evidence of millisecond spin periods of NSs in
about one-third of known Galactic LMXBs.  These measurements are
summarized in Fig.~\ref{fig:spin-histogram}a.  Altogether, there are
seven such NSs with firmly established spin periods, by either
pulsations in the persistent emission (discovered by Wijnands \& van
der Klis in the millisecond X-ray pulsar SAX~J1808.4-3658;
\cite{wijnands98}) or oscillations during type I X-ray bursts (burst
QPOs, first discovered in 4U~1728--34 by Strohmayer et
al. \cite{strohmayer96:_millis_x_ray_variab_accret}).  There are an
additional thirteen sources with twin kHz QPOs for which the spin may
be approximately equal to the frequency difference
\cite{klis99:_millis}.  A striking feature of all these neutron stars
is that their spin frequencies lie within a narrow range, $260\
\Hz<\nuspin<589\ \Hz$.  The frequency range might be even narrower if
the burst QPOs seen in KS~1731--260, MXB~1743--29, and Aql~X-1 are at
the first harmonic of the spin frequency, as is the case with the
$581\ \Hz$ burst oscillations in 4U~1636--536
\cite{miller99:_eviden_antip_hot_spots_durin}.  These NSs accrete at
diverse rates, from $10^{-11}\Mspyr$ to the Eddington limit,
$\Mdot_{\rm Edd}=2\times10^{-8}\Mspyr$. Since disk accretion exerts a
substantial torque on the NS and these systems are very old
\cite{vanParadijs95:LMXB_distrib}, it is remarkable that their spin
frequencies are so similar, and that none of them are near the breakup
frequency of $\approx1.5\ \kHz$.

One possible explanation, proposed by White \& Zhang \cite{white97},
is that these stars have reached the magnetic spin equilibrium (where
the spin frequency matches the Keplerian frequency at the
magnetosphere) at nearly identical frequencies.  This requires that
the NS dipolar $B$ field correlate very well with $\dot{M}$
\cite{white97,miller98}.  However, there are no direct $B$ field
measurements for LMXBs, and in the strongly magnetic binaries, where
the $B$ field {\it has} been measured directly, such a correlation is
not observed. More importantly, for 19 out of 20 systems, there must
be a way of hiding persistent pulses typically seen from magnetic
accretors.  These difficulties led Bildsten \cite{Bildsten98:GWs} to
resurrect the conjecture originally due to Papaloizou \& Pringle
\cite{Papaloizou78:gravity_waves} and Wagoner \cite{Wagoner84} that
gravitational radiation can balance the torque due to accretion. The
detailed mechanisms will be discussed in the following sections.

Regardless of the detailed mechanism for GW emission, if gravitational
radiation balances the accretion torque, then it is easy to estimate
the GW strength. As noted by Wagoner \cite{Wagoner84}, in equilibrium
the luminosities in GWs and in X-rays are both proportional to the
mass accretion rate $\dot{M}$, so the characteristic strain amplitude
$h_c$ depends on the X-ray flux $F_x$ at Earth and the spin frequency
\begin{equation}
h_c = 4 \times 10^{-27} {R_6^{3/4}\over M_{1.4}^{1/4}}
\Biggl(\frac{300 \ {\rm Hz}}{\nu_s}\Biggr)^{1/2}
\left(F_x\over 10^{-8} \ 
{\rm erg \ cm^{-2} \ s^{-1}}\right)^{1/2}.
\end{equation}
In Fig.~\ref{fig:ligo2}b we show $h_c$ for Sco~X-1 (marked with a
star), a few other bright LMXBs with the spin inferred from kHz QPO
separation (thick dots) and burst QPO frequency (triangles), and the
millisecond X-ray pulsar SAX~J1808.4-3658 (open diamond).  The dotted
line shows LIGO-II sensitivity $h_{3/{\rm yr}}$ (i.e., $h_c$
detectable with 99\% confidence in $10^7$~s, provided the frequency
and the phase of the signal are known in advance
\cite{bradycreighton99}) in the broadband configuration, while the
solid line shows $h_{3/{\rm yr}}$ for the narrowband configuration.
However, the frequency and the phase are known precisely only for the
SAX~J1808.4-3658 millisecond X-ray pulsar \cite{chakrabarty98b}.  For
other sources, Brady \& Creighton \cite{bradycreighton99} showed that
the number of trials needed to guess the poorly known orbital
parameters or to account for the torque noise due to $\dot M$
variations lowers the effective sensitivity by roughly a factor of
two.

While the average $\dot M$ certainly correlates with the X-ray
brightness, current observations unfortunately do not let us robustly
infer the {\it instantaneous\/} torque \cite{klis99:_millis}.  Even
though $\dot M$ varies on a timescale of days, torque noise leads to
frequency drift only on a timescale of weeks.  The accretion torque is
$N_a=\dot{M}(GMR)^{1/2}$, and the total time-averaged torque is zero
due to equilibrium with GW emission.  Assume that $N_a$ flips sign
randomly on a timescale $t_s\approx$~few days.  The spin frequency
$\Omega$ will experience a random walk with step size
$\delta\Omega=(N_a/I) t_s$, where $I$ is the NS moment of inertia.
After an observation time $t_{\rm obs}$, the drift is
$\Delta\Omega=(t_{\rm obs}/t_s)^{1/2}\delta\Omega$.  This will exceed
a Fourier frequency bin width, i.e., $\Delta\Omega\gtrsim2\pi/t_{\rm
obs}$ only after
\begin{equation}
t_{\rm obs}= 
\frac{21\ {\rm days}}{M_{1.4}^{1/3} R_6^{1/3}}
\left(\frac{1\ {\rm day}}{t_s}\right)^{1/3}
\left(\frac{10^{-8}\ \Mspyr}{\dot{M}}\right)^{2/3}.
\end{equation}
Hence, on a timescale of tens of days, the intrinsic GW signal is
coherent.  The dashed line in Fig.~\ref{fig:ligo2}b shows the LIGO-II
sensitivity for a two-week integration in a narrowband configuration.
This suggests that the way to detect GWs from LMXBs may be
short integrations \cite{bradycreighton99}.

Currently, there are two classes of theories for GW emission from NSs
in LMXBs.  The presence of a large-scale temperature asymmetry in the
deep crust will cause it to deform \cite{Bildsten98:GWs}. The
resulting ``mountains'' will give the rotating star a time-dependent
mass quadrupole moment.  Alternatively, unstable r-mode pulsations (see a
review by B.~Owen in this volume) of a suitable amplitude in the NS
liquid core can emit enough gravitational radiation to balance the
accretion torque \cite{Bildsten98:GWs,andersson99:accreting_rmode}.

\section{Deformations of Accreting NS Crusts}
\label{sec:crusts}

\begin{figure}[t]
\begin{center}
\epsfig{file=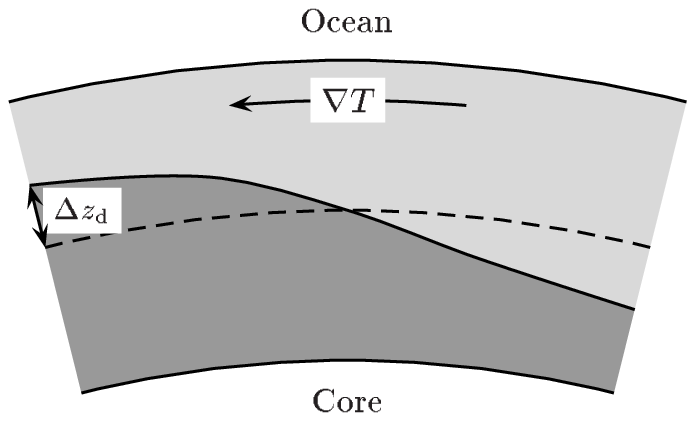,width=2.85in}
\epsfig{file=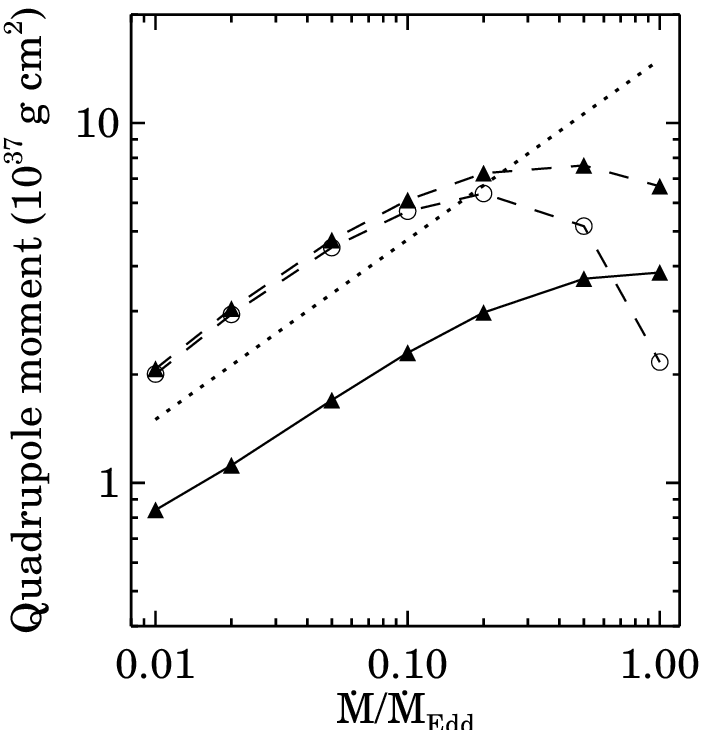,width=2.85in}
\end{center}
\caption[c]{\label{fig:sink-diagram}
Left (a) A cartoon description of how a transverse temperature
gradient leads to a varying altitude for the e$^-$ captures.
\label{fig:quadrup-mdot}
Right (b) The quadrupole $Q_{22}$ due to a single capture layer in the
inner crust as a function of $\dot{M}$. The solid and dashed lines
denote the results for several NS models for a $10\%$ composition
asymmetry, while the dotted line is the relation given by
Eq.~(\ref{eq:qneed}), i.e., the quadrupole necessary for spin
equilibrium at $\nu_s=300$~Hz as a function of $\dot{M}$.}
\end{figure}

The crust is a $\approx1$~km layer of crystalline ``ordinary'' (albeit
neutron-rich) matter that overlies the liquid core composed of free
neutrons, protons, and electrons.  The crust's composition varies with
depth in a rather abrupt manner.  As an accreted nucleus gets buried
under an increasingly thick layer of more recently accreted material,
it undergoes a series of e$^-$ captures, neutron emissions, and
pycnonuclear reactions \cite{Sato79,HZ90,Blaes90}, resulting in
layered composition.  In Fig.~\ref{fig:sink-diagram}a, we show
schematically two such compositional layers (light and dark shading)
sandwiched between the liquid core and the ocean.  Since an
appreciable fraction of the pressure is supplied by degenerate
electrons, e$^-$ captures induce abrupt density increases.  In the
outer crust, these density jumps are as large as $\approx10\%$, while
in the inner crust the density contrast is $\lesssim1\%$.  At $T=0$,
the e$^-$ captures occur when the electron Fermi energy $\Ef$ is
greater than the mass difference between the e$^-$ capturer and the
product of the reaction. In the absence of other effects, this depth
is the same everywhere, and such an axisymmetric capture boundary (the
dashed line in Fig.~\ref{fig:sink-diagram}a) does not create a mass
quadrupole moment.

However, in accreting NSs the crustal temperatures are high enough (in
excess of $2\times 10^8\ {\rm K}$) that e$^-$ capture rates become
temperature-sensitive \cite{BC98}. Bildsten \cite{Bildsten98:GWs}
pointed out that if there is a lateral temperature gradient in the
crust (the arrow in Fig.~\ref{fig:sink-diagram}a), then regions of the
crust that are hotter undergo captures at a lower density than the
colder regions.  The capture boundary becomes ``wavy'' (the solid line
in Fig.~\ref{fig:sink-diagram}a), with captures proceeding a height
$\dzd$ higher on the hot side of the star, and $\dzd$ lower on the
cold side. Such a temperature gradient, if misaligned from the spin
axis, will give rise to a nonaxisymmetric density variation and a
nonzero quadrupole moment $Q_{22}$ \cite{Bildsten98:GWs}.

The required quadrupole moment $Q_{\rm eq}$ such that GW emission is
in equilibrium with the accretion torque is
\begin{equation}\label{eq:qneed} 
Q_{\rm eq}=3.5\times 10^{37} {\rm g \ cm^2}M_{1.4}^{1/4}R_6^{1/4}
\left(\dot M\over 10^{-9} M_\odot {\rm \
yr^{-1}}\right)^{1/2}\Bigg(\frac{300 \ {\rm Hz}} {\nu_s}\Bigg)^{5/2},
\end{equation}
The range of $\dot M$'s in LMXBs is $\approx10^{-11}-2\times 10^{-8}\
\Msun\ {\rm yr^{-1}}$, requiring $Q_{22}\approx
10^{37}-10^{38}$~g~cm$^2$ for $\nu_s=300 \ {\rm Hz}$
\cite{Bildsten98:GWs}.  Can temperature-sensitive e$^-$ captures
sustain a quadrupole moment this large?  The quadrupole moment
generated by a temperature-sensitive capture boundary is $Q_{22}\sim
Q_{\rm fid} \equiv \Delta\rho\dzd R^4$, where $\Delta\rho$ is the
density jump at the electron capture interface. $Q_{\rm fid}$ is the
quadrupole moment that would result if the crust did not elastically
adjust (or just moved horizontally) in response to the lateral
pressure gradient due to wavy e$^-$ captures.  Using this estimate,
Bildsten \cite{Bildsten98:GWs} argued that a single wavy capture
boundary in the thin outer crust could generate $Q_{22}$ sufficient to
buffer the spinup due to accretion (Eq.~[\ref{eq:qneed}]), provided
that temperature variations of $\approx20\%$ are present in the crust.

 However an important piece of physics is missing from this estimate:
the shear modulus $\mu$.  If $\mu=0$, the crust becomes a liquid and
cannot support a non-zero $Q_{22}$.  Ushomirsky, Cutler, \& Bildsten
\cite{UCB99} recently calculated of the elastic response of the crust
to the wavy e$^-$ captures.  They found that the predominant response
of the crust to a lateral density perturbation is to sink, rather than
move sideways. For this reason, $Q_{22}$ generated in the {\it
outer\/} crust \cite{Bildsten98:GWs} is much too small to buffer the
accretion torque. However, a single e$^-$ capture boundary in the deep
{\it inner\/} crust can easily generate an adequate $Q_{22}$.  Because
of the much larger mass involved by captures in the inner crust, the
temperature contrasts required are $\lesssim5\%$, or only
$\approx10^6-10^7$~K, not $\approx10^8$~K as originally postulated
\cite{Bildsten98:GWs}.

What causes the lateral temperature asymmetry, and can it persist
despite the strong thermal contact with the almost perfectly
conducting core?  In LMXBs, the crusts are composed of the compressed
products of nuclear burning of the accreted material. The exact
composition depends on the local accretion rate, which could have a
significant non-axisymmetric piece due to, e.g., the presence of a
weak $B$ field. Moreover, except in the highest accretion rate LMXBs,
nearly all of the nuclear burning occurs in type I X-ray bursts.
Burst QPOs (see Sec.~\ref{sec:observations}) provide conclusive
evidence that bursts themselves are not axisymmetric. Until the origin
of this symmetry breaking is clearly understood, it is plausible to
postulate that these burst asymmetries get imprinted into the crustal
composition.

Ushomirsky et al. \cite{UCB99} showed that such a non-uniform
composition leads directly to lateral temperature variations $\delta
T$.  Horizontal variations in the charge-to-mass ratio $Z^2/A$ (which
determines the crustal conductivity) and/or nuclear energy release
modulate the radial heat flux in the crust and set up a
nonaxisymmetric $\delta T$.  The $\delta T$'s required to induce a
$Q_{22}\approx Q_{\rm eq}$ can easily be maintained if there is a
$\approx10\%$ asymmetry in the nuclear heating or $Z^2/A$.  So long as
accretion continues, these $\delta T$'s persist despite the strong
thermal contact with the isothermal NS core. 

The e$^-$ capture $Q_{22}$ calculations \cite{UCB99} are summarized in
Fig.~\ref{fig:quadrup-mdot}b.  If the size of the heating or $Z^2/A$
asymmetry is a constant fixed fraction, then for
$\dot{M}\lesssim0.5\dot{M}_{\rm Edd}$ the {\it scaling\/} of
$Q_{22}(\dot M)$ is just that needed for all of these NSs to have the
same spin frequency (the normalization is proportional to the
magnitude of the asymmetry, but the scaling is fixed by the
microphysics).  For $\Mdot\gtrsim0.5\Mdot_{\rm Edd}$, in order to
explain the spin clustering at {\it exactly\/} 300~Hz, this mechanism
requires that the crustal asymmetry correlate with $\Mdot$.
Alternatively, if the asymmetry is the same as in the low $\dot{M}$
systems, then one would expect the bright LMXBs to have higher spins,
a possibility that cannot be ruled out by current observations
(Sec.~\ref{sec:observations}).

So long as crustal deformations are due to shear forces only, the crustal
$Q_{22}$ is limited by the yield strain $\bar\sigma_{\rm max}$ to be
less than \cite{UCB99}
\begin{equation}
Q_{\rm max} \approx 10^{38}\ {\rm g\  cm}^2 
\left(\frac{\bar\sigma_{\rm max}}{10^{-2}}\right)
\frac{R_6^{6.26}}{M_{1.4}^{1.2}}.
\end{equation}
$Q_{22}$'s needed to buffer the accretion torque require strains
$\bar\sigma\approx10^{-3}-10^{-2}$ at $\approx300$~Hz, with
$\bar\sigma\gtrsim10^{-2}$ in near-Eddington accretors.  Estimates for
the yield strain of the neutron star crust range anywhere from
$10^{-1}$ for perfect one-component crystals to $10^{-5}$. Hence
$\bar\sigma\gtrsim10^{-2}$ is probably higher than yield strain,
though this conclusion is based on extrapolating experimental results
for terrestrial materials by $>10$ orders of magnitude.  Such high
strains are perhaps the biggest problem with the crustal $Q_{22}$
mechanism. At high pressures ($\gg$ shear modulus) terrestrial
materials tend to deform plastically rather than crack, and so the
crusts of accreting NSs may be in a state of continual plastic flow.
If accretion continually drives the crust to $\bar\sigma_{\rm max}$,
this leads to a natural explanation for spin similarities near
$\Mdot_{\rm Edd}$.

However, many fundamental issues remain unanswered. First, the
calculation \cite{UCB99} is only good up to an overall prefactor set
by the density of capture layers in the deep crust. We thus need an
exploratory calculation of both the composition of the products of
nuclear burning in the upper atmosphere over the entire range of
$\Mdot$ in LMXBs, and their detailed nuclear evolution under
compression in the crust. Knowledge of the composition is also
necessary for a robust calculation of the shear modulus, which is
clearly the crucial number to know when computing the elastic response
of the crust.  Recent results
\cite{PethickRavenhall95,pethick98:liquid_crystal} indicate that inner
crusts of NSs are composed of highly nonspherical nuclei and may be
more like liquid crystals (solids that provide no elastic restoring
force for certain kinds of distortions) rather than simple Coulomb
solids.  Such improved calculations have implications far beyond the
problem of the crustal quadrupole moment.  The shear modulus of the
crust affects the maximum elastic energy that can be stored in the
crust, and hence the energetics of pulsar glitches and starquakes, as
well as the models of magnetic field evolution that depend on crustal
``plate tectonics'' (see \cite{Ruderman91}).  It even has bearing on
the stability of r-modes in neutron stars (Sec.~\ref{sec:r-modes}). In
addition, much work needs to be done on understanding what sets the
shear strength $\bar\sigma_{\rm max}$ of multicomponent crystals,
likely with defects and highly nonspherical nuclei, or what happens
when $\bar\sigma_{\rm max}$ is exceeded and viscoelastic flow ensues.

\section{R-modes in Accreting NS Cores}
\label{sec:r-modes}

\begin{figure}[t]
\begin{center}
\epsfig{file=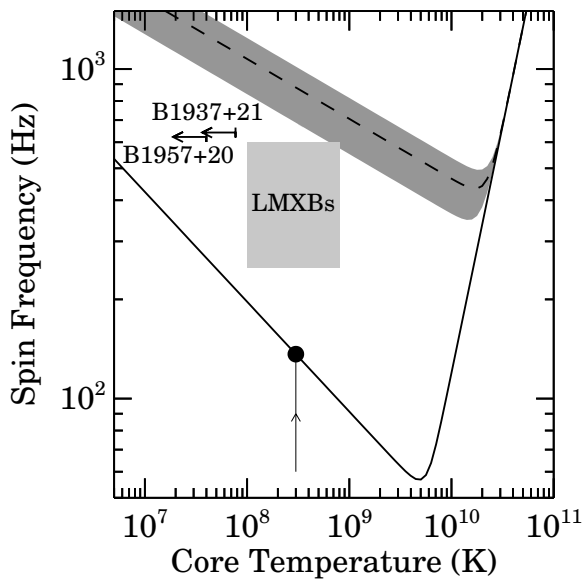}
\end{center}
\caption[c]{\label{fig:instability} Critical spin frequencies of the
r-mode instability.  The solid line is the critical frequency set only
by shear and bulk viscosities in the core (same as Fig.~1 of B.~Owen's
review in this volume).  The dashed line also includes viscous 
boundary layer damping, while the shading around it displays the
effect of core superfluidity.}
\end{figure}

Bildsten \cite{Bildsten98:GWs} and Andersson, Kokkotas, \& Stergioulas
\cite{andersson99:accreting_rmode} pointed out that the r-mode
instability (see the review by B.~Owen in this volume for the
introduction and notation) may also explain the spins of NSs in LMXBs,
and, if so, produce GW signal detectable by LIGO-II.  An accreting NS
is spun up (along a line in $(\nu_s,T)$ plane marked with an arrow in
Fig.~\ref{fig:instability}) until it reaches the r-mode instability
line (the solid line in Fig.~\ref{fig:instability}).  At that point
(marked by a thick dot in Fig.~\ref{fig:instability}) the r-mode
amplitude needed to balance the accretion torque is rather small.  The
NS can then hover at the instability line, with $1/\tau_G+1/\tau_V=0$,
and the r-mode amplitude such that it balances the accretion torque.
However, at $T=$~few$\times10^8$~K, the r-mode$-$accretion equilibrium
spin frequency would be $\approx150$~Hz, rather than $\approx300$~Hz,
resulting in an apparent disagreement with the observed spins of
LMXBs.  Bildsten \cite{Bildsten98:GWs} and Andersson et
al. \cite{andersson99:accreting_rmode} speculated that including other
sources of viscosity, e.g., superfluid mutual friction, is likely to
raise the instability curve, resulting in equilibrium frequencies
closer to the canonical $300$~Hz.  Finally, the narrow range of the
observed spin frequencies would presumably arise because of the
similar core temperatures of the accreting NSs (shown by the shaded
box in Fig.~\ref{fig:instability}).

Recent theoretical work brought up several challenges to this
scenario.  Levin \cite{Levin99} and Spruit \cite{spruit99:_gamma_x}
showed that steady-state equilibrium between accretion and r-modes is
thermally unstable for normal fluid cores.  In a normal fluid (i.e.,
not superfluid), the shear viscosity scales as $T^{-2}$, so the
increase in the core temperature due to viscous heating decreases the
shear viscosity. The smaller shear viscosity increases the growth rate
of the r-mode, leading to an unstable runaway. Using a
phenomenological model of nonlinear r-mode evolution
\cite{owen98:_gravit}, Levin \cite{Levin99} showed that in this case,
instead of just hovering near the instability line, the r-mode grows
rapidly until saturation, heats up the star, and spins it down and out
of the instability region in less than 1~yr.  Therefore, if NSs in
LMXBs have normal fluid cores, we would not expect to see any of them
with $\approx300$~Hz spins.

The unstable regime for r-modes in normal fluid NSs (above the solid
line in Fig.~\ref{fig:instability}) encompasses much of the parameter
space occupied by NSs in LMXBs and newborn NSs.  Because of the large
torques exerted by the unstable r-modes, we would not expect to see
any NSs in this region.  In addition, the existence of two 1.6~ms
radio pulsars (the spins and upper limits on core temperatures of
which are shown by arrows in Fig.~\ref{fig:instability}) means that
rapidly rotating NSs are formed in spite of the r-mode instability
\cite{andersson99:accreting_rmode}. While it is not clear whether
their {\it current\/} core temperatures place these pulsars within the
r-mode instability region, normal-fluid r-mode theory says that they
were certainly unstable during spinup.

{\it Superfluid\/} r-mode calculations have been eagerly awaited, as
they could resolve these conflicts.  However, Lindblom \& Mendell
\cite{lindblom99:superfluid} showed that, for most values of the
neutron-proton entrainment parameter, the superfluid dissipation is
not competitive with gravitational radiation.  Only over about $3\%$
of the possible entrainment parameter values is mutual friction strong
enough to compete with gravitational radiation.  The r-mode
instability line in this case is an approximately horizontal line (see
Fig.~8 of \cite{lindblom99:superfluid}) separating the unstable spin
frequencies ($\nu_s > \nu_{\rm crit}$) from the stable ones ($\nu_s <
\nu_{\rm crit}$). If the superfluid entrainment parameter has a value
such that $\nu_{\rm crit}\approx300$~Hz, then the LMXB spin
frequencies could still be understood in terms of the r-mode
instability and the special nature of the NS superfluid.

Before learning about these results, Brown \& Ushomirsky
\cite{bu99:rmodes} ruled out such a simple superfluid equilibrium
{\it observationally\/} for a subset of LMXBs.  In steady state, the
shear in the r-mode deposits $\approx10$~MeV of heat per accreted
baryon into the NS core.  When the core is superfluid, Urca neutrino
emission from it is suppressed, and this heat must flow to the NS
surface and be radiated thermally.  In steadily accreting systems
(such as Sco~X-1) this thermal emission is dwarfed by the accretion
luminosity of $GM\dot{M}/R\approx200$~MeV per accreted baryon.
However, in {\it transiently\/} accreting systems, such as Aql~X-1,
when accretion ceases, the r-mode heating should be directly
detectable as enhanced X-ray luminosity from the NS surface.  For
Aql~X-1 and other NS transients, Brown \& Ushomirsky
\cite{bu99:rmodes} showed that, if the superfluid r-mode equilibrium
prevails, then the quiescent luminosity should be about 5$-$10 times
greater than is actually observed.

A possible resolution of this conundrum has been recently proposed by
Bildsten \& Ushomirsky \cite{BU99}. All but the hottest
($\gtrsim10^{10}$~K) NSs have solid crusts. The r-mode
motions are mostly transverse, and reach their maximum amplitude near
the crust-core boundary. The fluid therefore rubs against the crust,
which creates a thin (few cm) boundary layer.  Because of the short
length scale, the dissipation in this boundary layer is very
large. The damping time due to rubbing is \cite{BU99}
\begin{equation}\label{eq:tau-rub}
\tau_{\rm{rub}}	\approx100{\rm~s~}T_8
	\frac{M_{1.4}}{R_6^2}
	\left(\frac{1{\rm~kHz}}{\nu_s}\right)^{1/2},
\end{equation}
substantially shorter than the viscous damping times due to the shear
and bulk viscosities in the stellar interior, as well as the mutual
friction damping time for most values of the superfluid entrainment
parameter.

The critical frequency for the r-mode instability in NSs with crusts
is shown by the dashed line in Fig.~\ref{fig:instability} for the case
where all nucleons are normal, and the dark shading around it
represents the range of frequencies when either neutrons or all
nucleons are superfluid.  The crust-core rubbing raises the minimum
frequency for the r-mode instability in NSs with crusts to
$\gtrsim500$~Hz for $T\approx10^{10}$~K, nearly a factor of five
higher than previous estimates.  This substantially reduces the
parameter space for the instability to operate, especially for older,
colder NSs, such as those accreting in binaries and millisecond
pulsars.  In particular, the smallest unstable frequency for the
temperatures characteristic of LMXBs is $\gtrsim700$~Hz, safely
above all measured spin frequencies. This work resolves the
discrepancy between the theoretical understanding of the r-mode
instability and the observations of millisecond pulsars and LMXBs,
and, along with observational inferences \cite{bu99:rmodes}, likely
rules out r-modes as the explanation for the clustering of spin
frequencies of neutron stars in LMXBs around 300~Hz.

To summarize, a significant role of {\it steady-state\/} r-modes in
LMXBs has probably been ruled out, both on theoretical grounds
\cite{BU99} (unless crust-core coupling is much stronger than was
estimated), and observationally \cite{bu99:rmodes} (for Aql~X-1 in
particular).  However, stochastically excited r-modes that decay
rapidly may still play a significant role in accreting systems, as
even a very small amplitude ($\alpha\lesssim10^{-5}$) can balance the
accretion torque at $\Mdot_{\rm Edd}$.  In addition, the issues of
crustal shear modulus and the structure of the crust-core boundary,
highlighted in Sec.~\ref{sec:crusts}, are of paramount importance for
r-modes as well. Crustal quadrupoles \cite{Bildsten98:GWs,UCB99} can
explain the spins of LMXBs and remain a viable source of continuous
GWs, but the strains at $\Mdot_{\rm Edd}$ are rather high.  The
crustal breaking strain $\bar\sigma_{\rm max}$ is not likely to be
understood theoretically any time soon, and detection of GWs from LMXBs
with LIGO-II type instruments will surely teach us many new things about
NSs.

\acknowledgments GU acknowledges support from the Fannie and John
Hertz Foundation.

\end{document}